\documentstyle[psfig]{article}
\begin{document}
\title{Stoppage of Light Made Flexible by an Additional Control Field}
\author{G. S. Agarwal and Tarak Nath Dey\\\large {Physical Research Laboratory,
Navrangpura}\\\large{ Ahmedabad 380 009, Gujarat, India}}

\date{\today}
\maketitle
\begin{abstract}
We show how the application of a coupling field connecting the two lower
metastable states of a Lambda system facilitates  stoppage of light in a coherently
driven Doppler broadened atomic medium via electromagnetically induced transparency.
\end{abstract}

\newpage
In recent times one has learnt how the propagation of electromagnetic pulse
could be manipulated by the use of the appropriate coherent
fields$[1-13]$.
 This is because the propagation is dependent on the dispersive properties
\cite{Gar} of the medium and it is now understood how the dispersion of a
medium can be manipulated by using control fields\cite{Harr,Tewari}. As a
matter of fact, regions of anomalous dispersion could be converted into
regions of normal dispersion. It has been shown both theoretically and
experimentally how the control fields could produce spatially
 varying refractive index profiles\cite{Kapoor}. All
these have remarkable bearing on the pulse propagation. Harris and coworkers
$[1-2]$ first proposed how these control fields could be used to
produce slow light. This was followed by a series of experiments in Bose
condensates\cite{Hau} as well as in hot atomic vapors $[4-6]$.
 These experiments demonstrated the production of ultraslow light. In a
related development Wang {\it et al.} \cite{Wang} demonstrated the production
of superluminal propagation. They used the novel idea of using a
bichromatic pump to produce a region of anomalous dispersion with negligible
absorption or gain. A recent paper shows how the pulse propagation could be
subjected to a new control parameter so that the same system can exhibit both
sub- and superluminal propagation\cite{Agarwal}. The experiments of Budker
{{\it et al.}\cite{Bud} also exhibit superluminal
behavior by the application of a suitable magnetic field. More recent experiments have demonstrated how the light
could be stored in atomic coherences\cite{Liu,Phillips} and how it can be
retrieved at will. All these depend on the temporal dispersion of the medium
and the experiments are interpreted in terms of dark polaritons
\cite{Fleischhauer}.

In a recent work Kocharovskaya {\it et al.}\cite{Kochar} proposed how the
spatial dispersion \cite{Agarnovich} can be used to stop light in a hot gas. They argued
that the motion of atoms leads naturally to a refractive index or a
susceptibility that is dependent on both the propagation vector and frequency.
Explicit calculation for a $\Lambda$ system shows that the stoppage of light
occurs when the control fields is suitably detuned from the atomic transition
and when the central frequency of the probe pulse satisfies the two-photon
resonance condition.The mechanism proposed by Kocharovskaya {\it et
al.}\cite{Kochar} is {\bf quite
different from the one of Ref}. \cite{Fleischhauer},as it does not require the
switching off and switching on of the control field.

In the present Communication we show how we can make the stoppage of light
flexible by using another control field. We consider a $\Lambda$ system as shown
in Fig. 1(a). We apply two control fields one on the transition $|1\rangle
\leftrightarrow |2\rangle$ and the other on the transition $|2\rangle
\leftrightarrow |3\rangle$. The probe pulse acts on the transition $|1\rangle
\leftrightarrow |3\rangle$. The transition $|2\rangle\leftrightarrow |3\rangle$
is generally electric dipole forbidden transition. The states $|2\rangle$ and
$|3\rangle$ are metastable states.
 We calculate the group velocity of the pulse
for different strengths of the two control fields. We demonstrate existence of very wide
region of parameters where the stoppage of light can be achieved while
maintaining regions of very low absorption.

In general for a spatially dispersive medium\cite{Agarnovich} the response of the medium
can be expressed by a
 susceptibility $\chi({\rm k},\omega)$ that is dependent on the propagation
 vector and frequency. Further the allowed wavevectors
are given by the dispersion relation
\begin{equation}
{\rm {k}^2}=\frac{{\omega}^2}{c^2}\left[1+4\pi\chi({\rm k},\omega)\right].
\end{equation}
It is well known that the above dispersion relation can lead to many real
solutions for k for a fixed $\omega$ and thus one has the possibility of additional waves in a
spatially dispersive medium. In this paper we however consider only the case
when $|\chi|$ is much smaller than one. In this case we would basically have one
wave. A simple analysis now shows that the group velocity is given by
\begin{equation}
v_g=\left[\frac{c\left(1-2\pi{\rm k}\frac{\partial \chi}{\partial
{\rm k}}\right)}{1+2\pi\omega\frac{\partial \chi}{\partial \omega}}\right].
\end{equation}
This formula assumes weak spatial as well as temporal dispersion.
 Besides  we assume that absorption is negligible. In a gas of
atoms, the $\chi({\rm k},\omega)$ is to be replaced by the average values
$\chi(\omega-{\rm k}v)$ over the distribution of velocities. Then the
expression for the group velocity becomes
\begin{equation}
v_g={\rm Re}\left[\frac{c\left(1+2\pi{\rm k}\langle v\frac{\partial \chi}{\partial
\omega}\rangle\right)}{1+2\pi\omega\langle\frac{\partial \chi}{\partial \omega}\rangle}
\right].
\end{equation}
Note that
\begin{equation}
\langle v \frac{\partial \chi}{\partial \omega}\rangle \neq \langle
v\rangle\langle\frac{\partial \chi}{\partial \omega}\rangle\neq0,
\end{equation}
and hence the stoppage of light takes place if the numerator in Eq. (3)
vanishes. The
susceptibility $\chi(\omega)$ will depend strongly on the intensities
and the frequencies of the two control fields. The susceptibility
$\chi(\omega)$ is obtained by solving the density matrix equations for
the $\Lambda$ system of Fig. 1(a), i.e, by calculating the density matrix element
$\rho_{_{13}}$ to first order in the applied optical field \cite{footnote} on the
transition$|1\rangle\leftrightarrow|3\rangle$ but to all orders in the two
control fields. We define all fields as
\begin{equation}
\vec{E}(z,t)=\vec{\mathcal {E}}(z)e^{-i(\omega t + \phi)} + {\textrm c.c.}
\end{equation}
By making a unitary transformation from
the density matrix $\rho$ to $\sigma$ via
\begin{equation}
\rho_{12}=\sigma_{12}e^{-i(\omega_2
t+\phi_2)},~\rho_{13}=\sigma_{13}e^{-i(\omega_2+\omega_3)t-i(\phi_2+\phi_3)},~\rho_{23}=\sigma
_{23}e^{-i(\omega_3 t+\phi_3)},
\end{equation}
we have the relevant density matrix equations
\begin{eqnarray}
\dot{\sigma}_{11}&=&i G \sigma_{21}+i g e^{-i(\Delta_4
t+\delta\phi)}\sigma_{31}-i G^{*} \sigma_{12}-i g^{*} e^{i(\Delta_4
t+\delta\phi)}\sigma_{13}-2(\gamma_1+\gamma_2)\sigma_{11}~,\nonumber\\
\dot{\sigma}_{22}&=& i G^* \sigma_{12} + i \Omega \sigma_{32} - i G \sigma_{21}
- i \Omega^* \sigma_{23} + 2\gamma_2\sigma_{11}~,\nonumber\\
\dot{\sigma}_{12}&=&-[\gamma_1+\gamma_2+\Gamma_{12}-i\Delta_2]\sigma_{12}+i G
\sigma_{22}+ig e^{-i(\Delta_4 t+\delta\phi)}\sigma_{32}\nonumber\\
&-&iG\sigma_{11}-i \Omega^*\sigma_{13}~,\\
\dot{\sigma}_{13}&=&-[\gamma_1+\gamma_2+\Gamma_{13}-i(\Delta_2+\Delta_3)]\sigma_{13}
+iG\sigma_{23}+ige^{-i(\Delta_4 t+\delta\phi)}\sigma_{33}\nonumber\\
&-&ige^{-i(\Delta_4t+\delta\phi)}\sigma_{11}-i\Omega\sigma_{12}~,\nonumber\\
\dot{\sigma}_{23}&=&-(\Gamma_{23}-i\Delta_3)\sigma_{23}+iG^*\sigma_{13}+i\Omega
\sigma_{33}-ige^{-i(\Delta_4 t+\delta\phi)}\sigma_{21}-i\Omega\sigma_{22}~.\nonumber
\end{eqnarray}
Here $\Gamma$'s give collisional dephasing terms; $\gamma$'s give the radiative
decay of the state $|1\rangle$;  $\Delta_i$'s are the detunings

\begin{equation}
\Delta_1=\omega_1-\omega_{13},~~\Delta_2=\omega_2-\omega_{12},~~\Delta_3
=\omega_3-\omega_{23},~~\Delta_4=\omega_1-\omega_2-\omega_3,
\end{equation}
and the relative phase difference of the probe field to the sum of the pump and
additional control field is
\begin{equation}
~~\delta\phi=\phi_1-
\phi_2-\phi_3.
\end{equation}
The coupling constants $2g$ and $2G$ are the Rabi frequencies of
the probe and the optical control field, while
$g=\vec{d}_{13}\cdot\vec{\mathcal{E}}_p/\hbar$ and
$G=\vec{d}_{12}\cdot\vec{\mathcal{E}}_c/\hbar$. The parameter
$\Omega$ characterizes the coupling between two lower levels. We
will refer to this field as LL coupling field. The susceptibility
$\chi$ can be obtained by considering the steady state solution of
(7) to first order in the probe field  on the transition $|1
\rangle \leftrightarrow |3 \rangle$. For this purpose we assume
$\gamma_1=\gamma_2=\gamma$ and write the solution as
\begin{equation}
\sigma=\sigma^0~+~\frac{g}{\gamma}~e^{-i(\Delta_4
t+\delta\phi)}~\sigma^+~+~\frac{g^*}{\gamma}~e^{i(\Delta_4 t+\delta\phi)}~\sigma^-+......
\end{equation}
On combining Eqs.(6) and (10), we note that to first order $\rho_{_{13}}(t)\equiv
(g/\gamma)$$e^{-i(\omega_1 t + \phi_1)}$\\$\sigma^{+}_{_{13}}$.
Thus $13$ element of $\sigma^+$ will give the susceptibility at the frequency
$\omega_1$ which now can be expressed in the form
\begin{equation}
\chi(\omega_1)=\frac{n|{\vec d}_{13}|^2}{\hbar \gamma}\sigma_{13}^+,
\end{equation}
where $n$ is the density of the atoms.
The phase dependence of different fields does not appear in the susceptibility .
Note that the zeroth order contribution in Eqs.(10) can result in  phase
dependent
components. However for the range of parameters used in this paper, the zeroth
order term is so small that it can be ignored.

To obtain the probe response in a Doppler-broadened medium $\sigma_{_{13}}^+$
should be averaged over the Maxwell-Boltzmann velocity distribution of the
moving atoms. For a single atom, moving with a velocity $v$ along the $z$ axis, the
probe frequency $\omega_1(v)$ and frequencies $\omega_2(v)$,
$\omega_3(v)$ of the two control fields as seen by the atom are given by
\begin{equation}
\omega_1(v)=\omega_1-{\rm k_1}v,~~\omega_2(v)=\omega_2-{\rm
k_2}v,~~\omega_3(v)=\omega_3-{\rm k_3}v.
\end{equation}
Thus susceptibilities for moving atoms are obtained by using the replacement
(12) in the solution of Eqs. (7). Note that the velocity dependence of $\omega_3$
is insignificant and can be dropped. For simplicity we can also set $k_1\approx
k_2$. These susceptibilities are to be averaged over the Maxwell-Boltzmann
distribution for the atomic velocities, defined by
\begin{equation}
{\rm P}(k_1v)d(k_1v)=\frac{1}{\sqrt{2\pi{\rm D^2}}}e^{-(k_1v)^2/2{\rm D}^2}d(k_1v),~~
{\rm D}=\sqrt{K_{B}T\omega_{1}^2 /M c^2}.
\end{equation}
We next give
the results of our calculations for the model systems shown in the Fig. 1(a). We
show a number of numerical results in Figs. 1, 2 and 3. In Fig. 1(b) and Fig. 1(c)
we show the behavior of the susceptibility as a function of the  detuning of the
probe when the control field $\omega_2$ is detuned, $\Delta_2=-50\gamma$.
It is clear from the Fig. 1(b) that with increase of the microwave field
intensity results in decrease of the probe absorption in presence of collisional
dephasing. At two-photon resonance condition i.e.,
$\Delta_1=\Delta_2=-50\gamma$, the absorption of the probe is very small.
Therefore the transparency window is obtained at
$\Delta_1=\Delta_2$. Note that the transparency dip that appears in the absorption
spectrum has finite bandwidth and its width depends on coupling fields $G$ and
$\Omega$. The probe pulse spectrum should be
contained within this finite bandwidth. However, the transparency dip is a
accompanied by a steep variation of $\langle {\rm Re}[\chi]\rangle$ with
probe detuning.
 We find that if two control fields are {\it suitably detuned} then the
light can be stopped.
We show in the Fig. 2(a) how the group velocity $v_g$ as defined in
Eqs. (3), changes from  negative values to large positive values as the
intensity of the LL coupling field is increased. The group velocity $v_g$ become
zero at the value of $\Omega=1741\times10^{-6}\gamma$ when the control field $G$
is out of one-photon resonance but satisfies the two photon resonance condition
($\Delta_1=\Delta_2=-50\gamma$). Note that for $^{87}{\rm Rb}$, a Rabi
frequency of $10^{-6}\gamma$ implies a magnetic field of the order of
 .993$\mu$G. The slope of
$\langle {\rm Re}[\chi]\rangle$ with respect to central frequency of the probe
pulse depends on the intensity of the two control fields and density of atoms.
The group velocity becomes zero [Fig. 2(b)] as the numerator in Eq. (3) changes sign
when the LL
coupling field is increased. The Fig. 2(c) gives the group velocity in the absence
 of spatial dispersion. A comparison of the Figs. 2(a) and 2(c) shows the important
role played by spatial dispersion. Further we notice from the Fig.
2(d) that at resonance condition~$(\Delta_1=\Delta_2=0)$ light cannot be
stopped. In order to understand how the application of the microwave field leads
to the stoppage of light, we show in the Fig. 3 the behavior of the Doppler
average of $\langle\partial \chi/\partial \omega_1\rangle$ and
$\langle1-2\pi {\rm k}_1\frac{\partial
\chi}{\partial \rm {k}_1}\rangle$ as
a function of $\omega_1$. The latter quantity crosses zero which results in the
stoppage of light. The Figs. 2 and 3 show how the application of the microwave
field changes all the physical quantities. At a more fundamental level the
microwave field and pump field together produce new dressed states of the
system. Such dressed states determine the response of the system to the applied
probe field. However all these can be understood easily only for a
homogeneously broadened system.

Thus in conclusion we have demonstrated how the application of an LL coupling
field in the $\Lambda$ system helps one to change the group velocity of
the pulse inside the medium from a negative to positive value, and thereby helps
in stopping light. Thus for a suitable detuning of
the pump and probe fields, one can stop light by just changing the intensity of the
LL coupling field.\\
G. S. Agarwal thanks M. O. Scully for useful discussions on the subject of pulse
propagation.

\newpage
\begin{figure}
\centerline{\begin{tabular}{c}
\psfig{figure=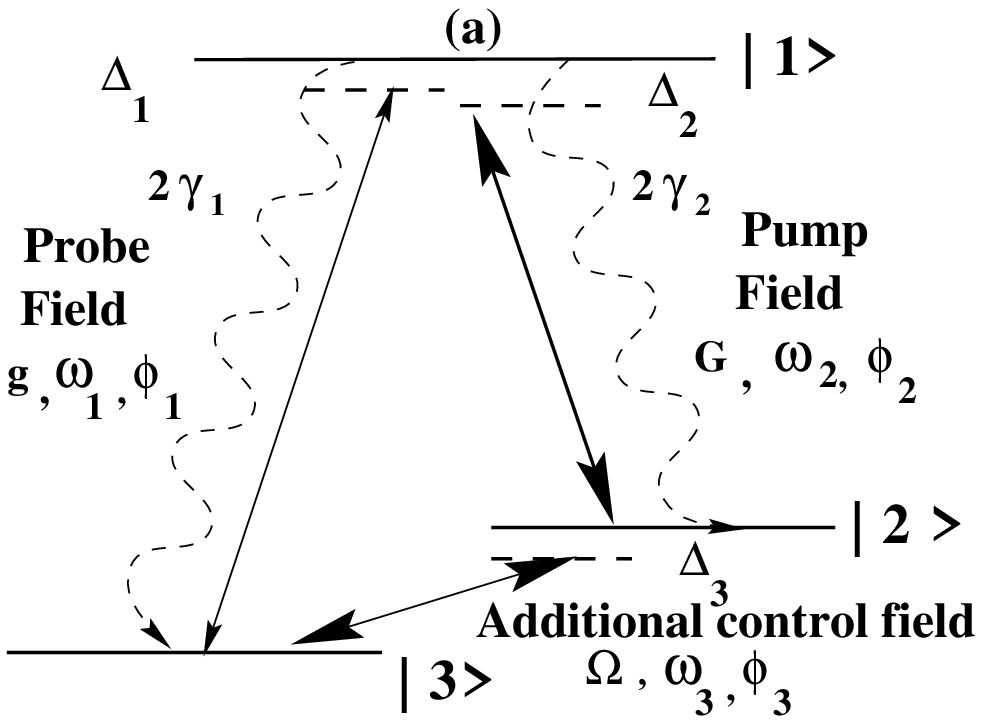,height=8.0cm,width=9.5cm}
\end{tabular}}
\end{figure}
\begin{figure}
\centerline{\begin{tabular}{cc}
\psfig{figure=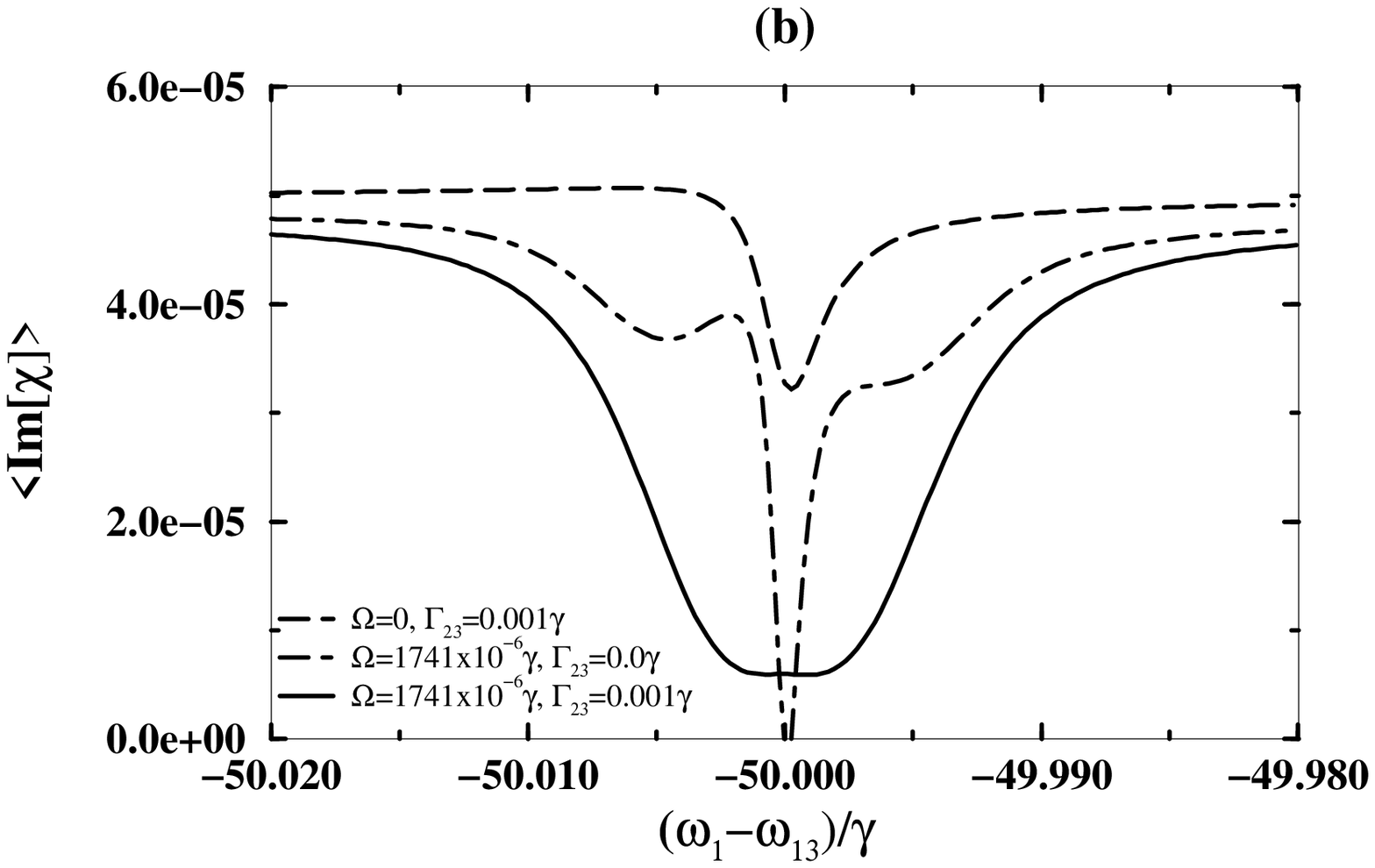,height=7.5cm,width=9.5cm}&
\psfig{figure=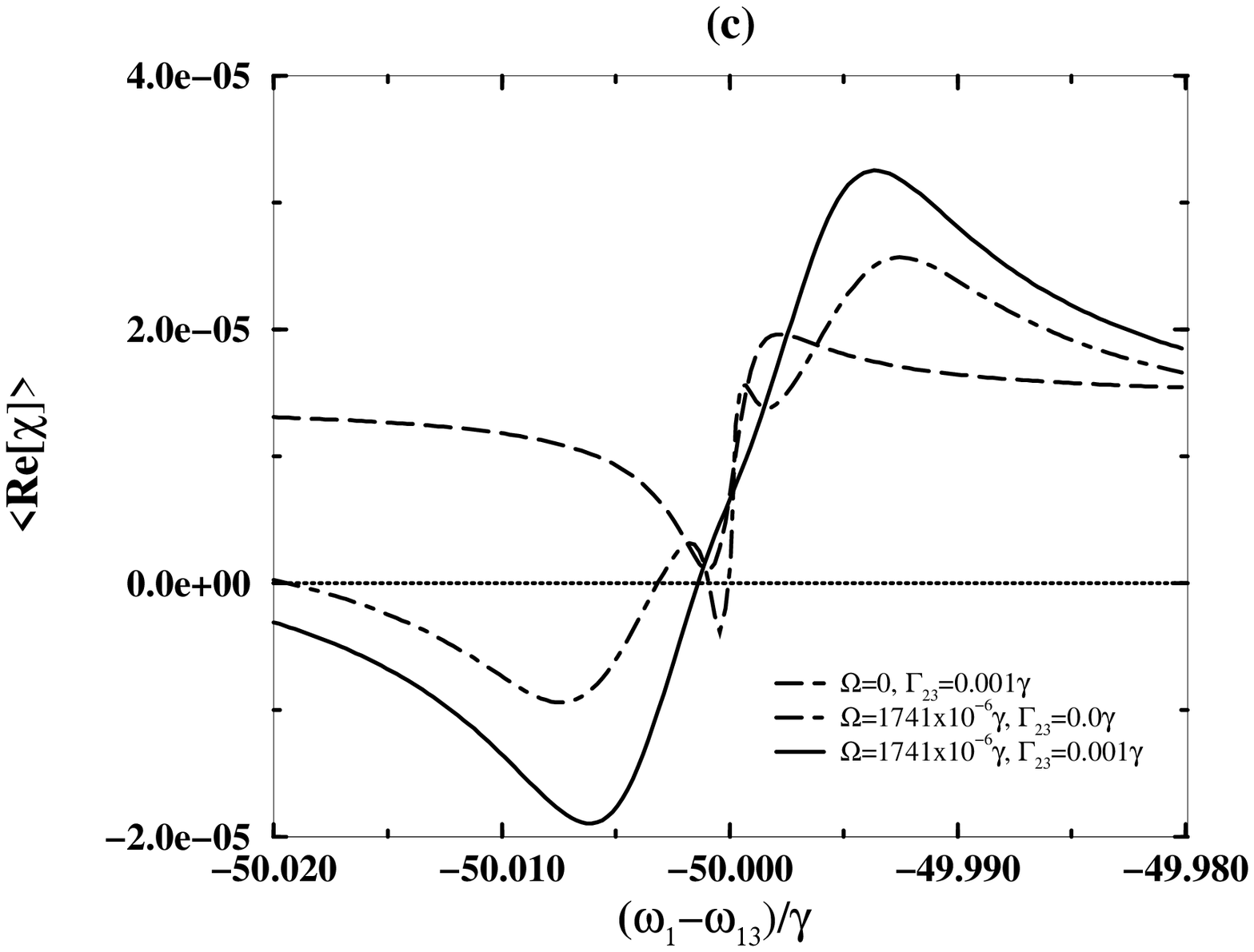,height=7.5cm,width=9.5cm}
\end{tabular}}
\caption{(a) Schematic diagram of three level $\Lambda$-system; the probe pulse
is applied on the transition $|1\rangle\leftrightarrow|3\rangle$; other fields
are cw.
(b) and (c) The imaginary
and real parts  of susceptibility $\langle[\chi]\rangle$at the
probe frequency $\omega_1$ in the presence of control field $G$ and LL coupling
field $\Omega$. Detuning $\Delta_2$ of the control field  is chosen as $-50\gamma$.
 The common
parameters of the above three graphs for $^{87}$Rb vapor are chosen as:
Doppler width parameter D$=1.33\times10^9$ rad/sec, density
$n=10^{12}$ atoms/cc, $G=0.3\gamma$, $\Delta_3=0$,
$\Gamma_{12}=\Gamma_{13}=0$,
$\gamma=3\pi\times10^{6}$~rad/sec.}
\end{figure}

\newpage
\begin{figure}
\centerline{\begin{tabular}{cc}
\psfig{figure=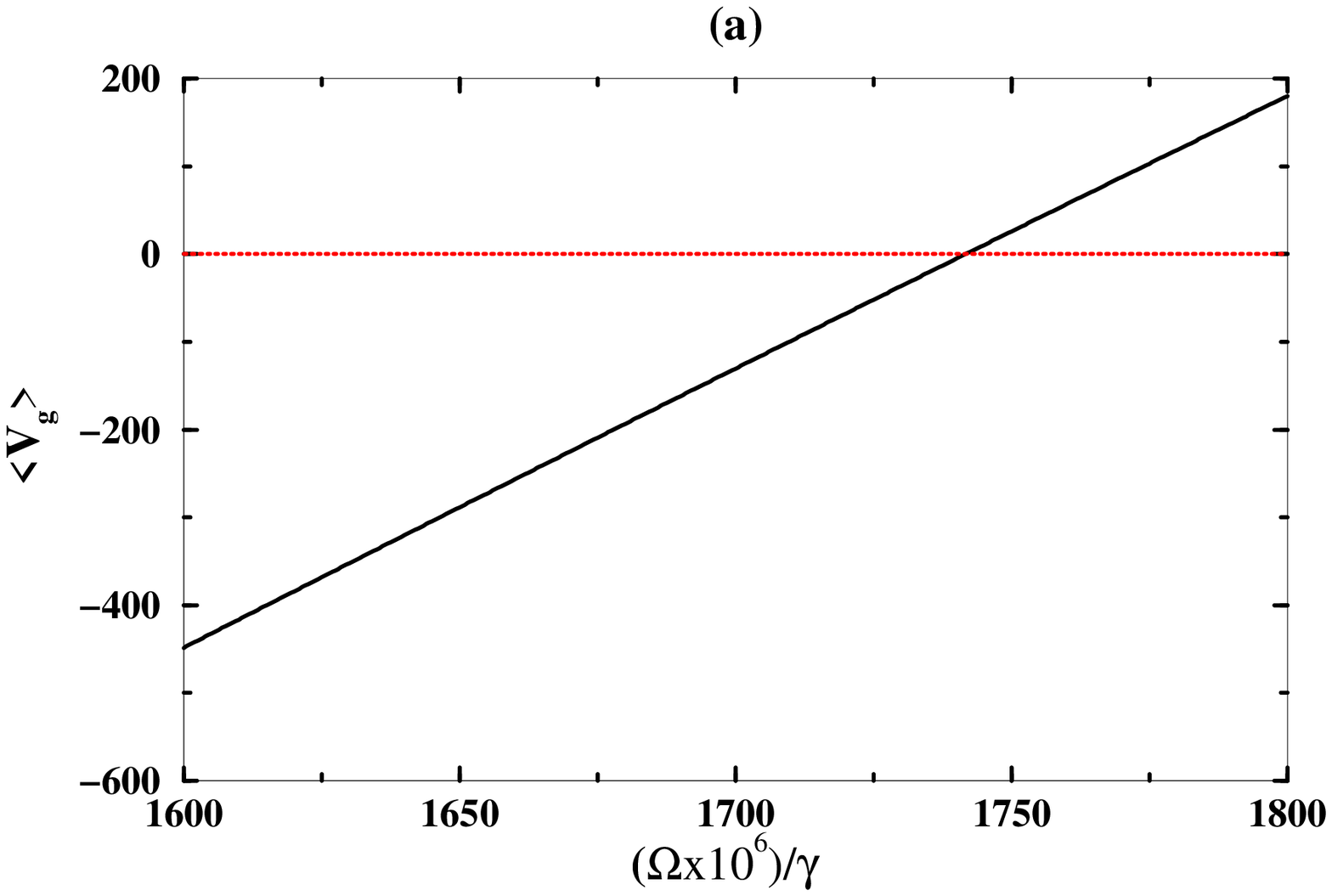,height=7.5cm,width=9.5cm}&
\psfig{figure=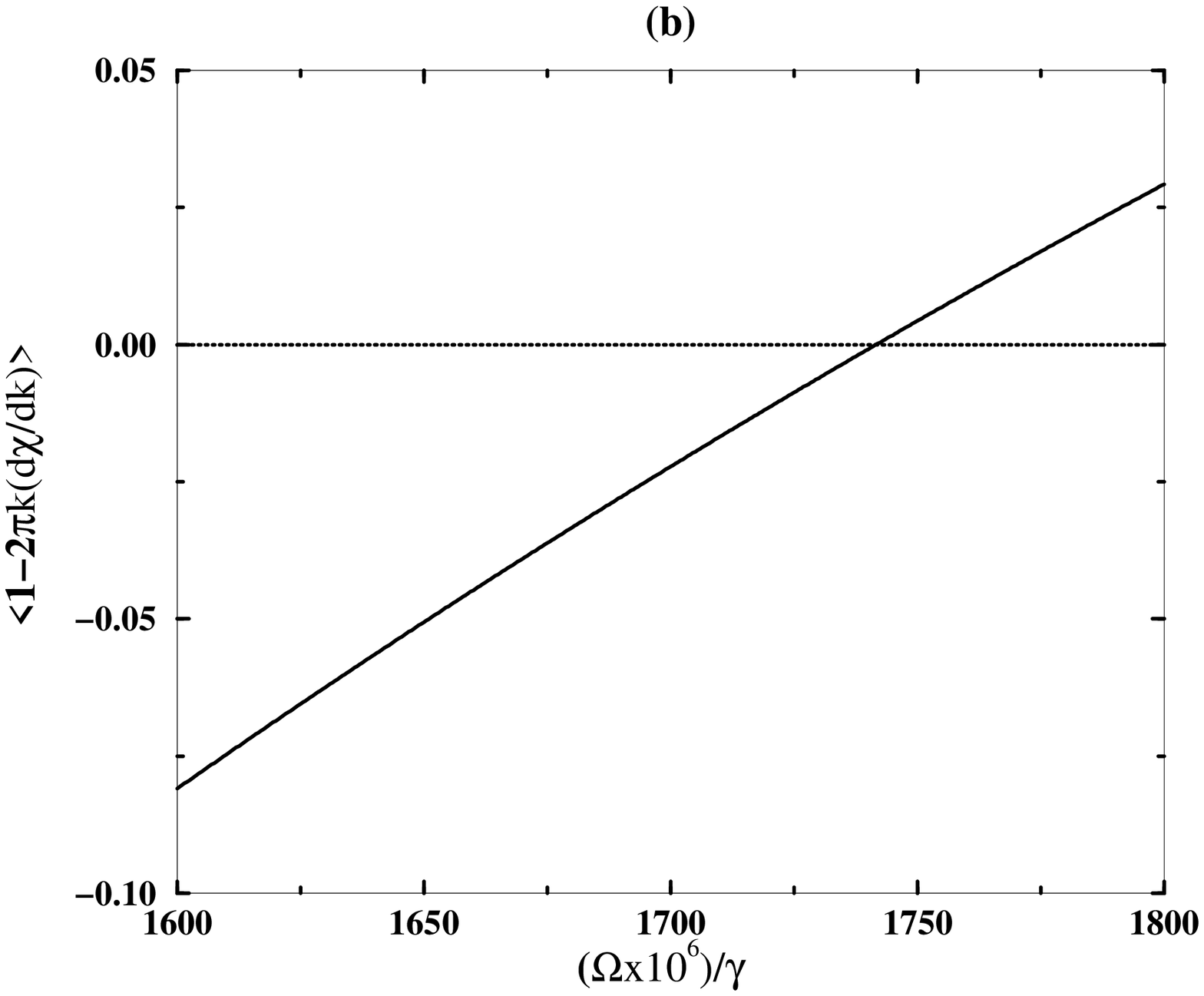,height=7.5cm,width=9.5cm}\\\\
\psfig{figure=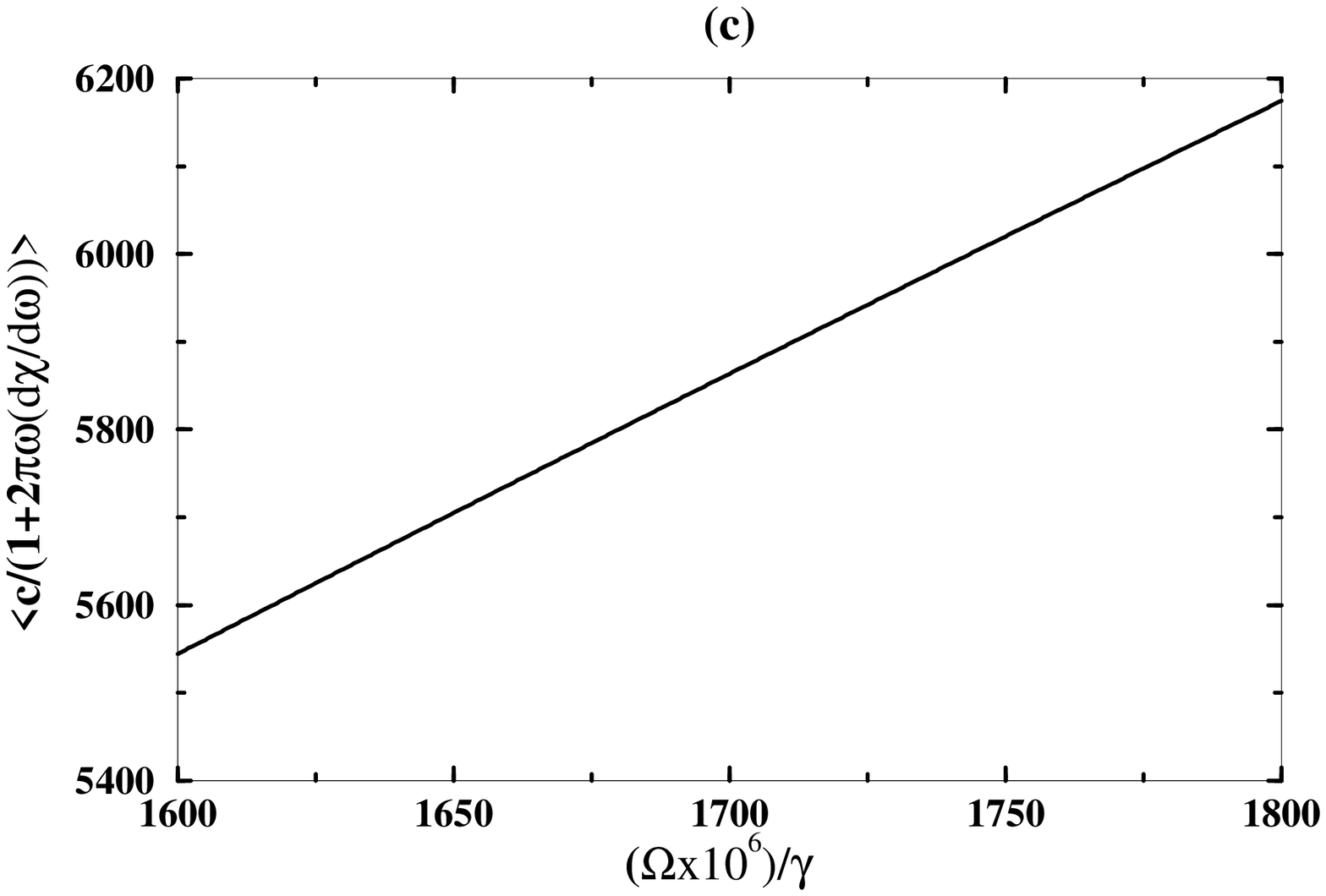,height=7.5cm,width=9.5cm}&
\psfig{figure=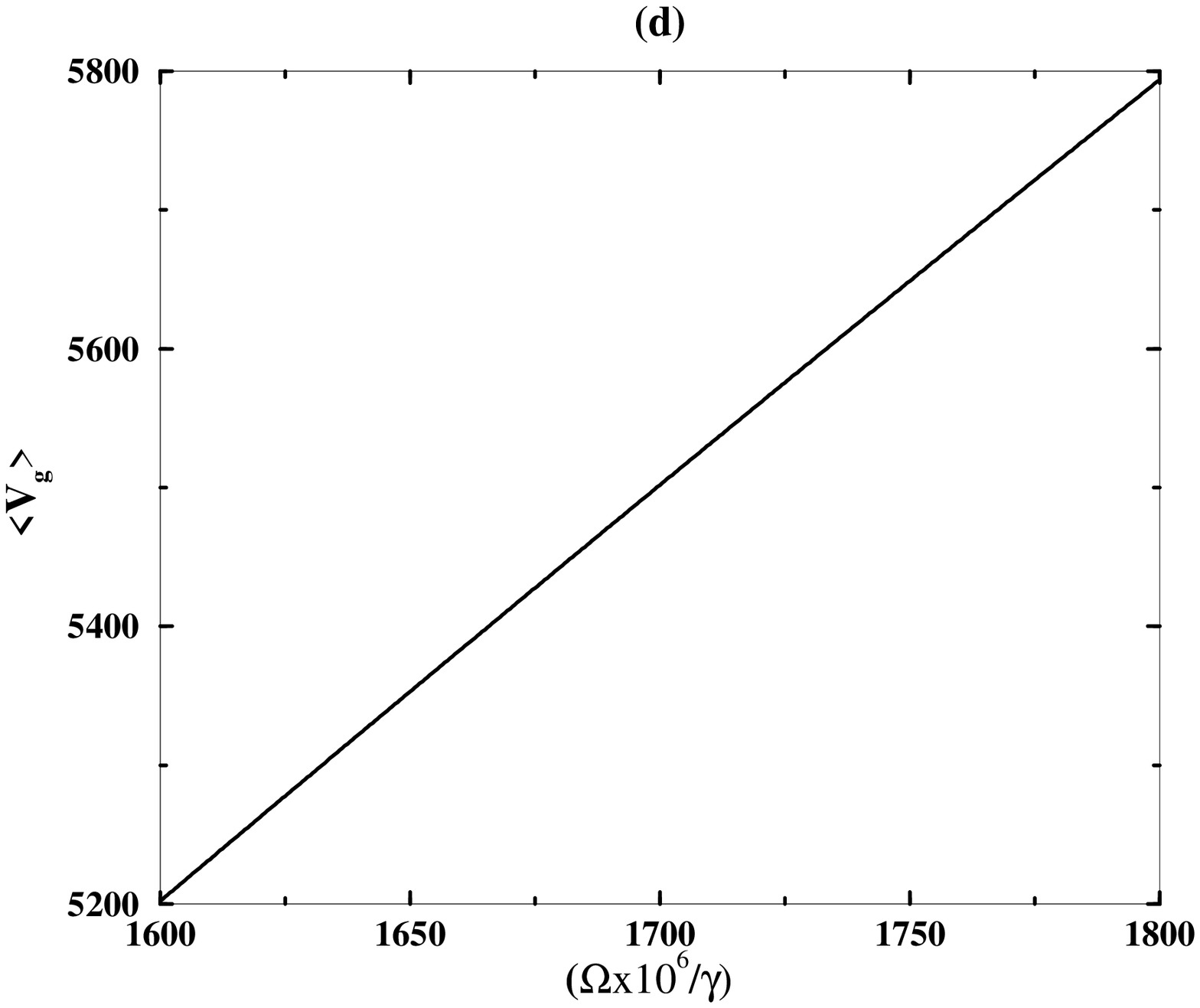,height=7.5cm,width=9.5cm}
\end{tabular}}
\caption{\small{
(a) shows variation of group velocity, in cm per sec, of Eq.(3) with the strength $\Omega$
of the LL coupling field. The group velocity becomes zero because the
numerator in Eq. (3) becomes zero as shown in the Fig. 2(b). The Fig. 2(c) gives
the behavior of the group velocity if the spatial dispersion of the
susceptibility were ignored.  The common
parameters of the above three graphs for $^{87}$Rb vapor are chosen as:
Doppler width parameter D$=1.33\times10^9$ rad/sec, density
$n=10^{12}$ atoms/cc, $G=0.3\gamma$, $\Delta_3=0$,
$\Gamma_{12}=\Gamma_{13}=0$, $\Gamma_{23}=0.001\gamma$,
$\gamma=3\pi\times10^{6}$~rad/sec, $\Delta_1=\Delta_2=-50\gamma$. For
comparison we also show in the Fig. 2(d) the
result for $\Delta_1=\Delta_2=0$.}}
\end{figure}

\begin{figure}
\centerline{\begin{tabular}{cc}
\psfig{figure=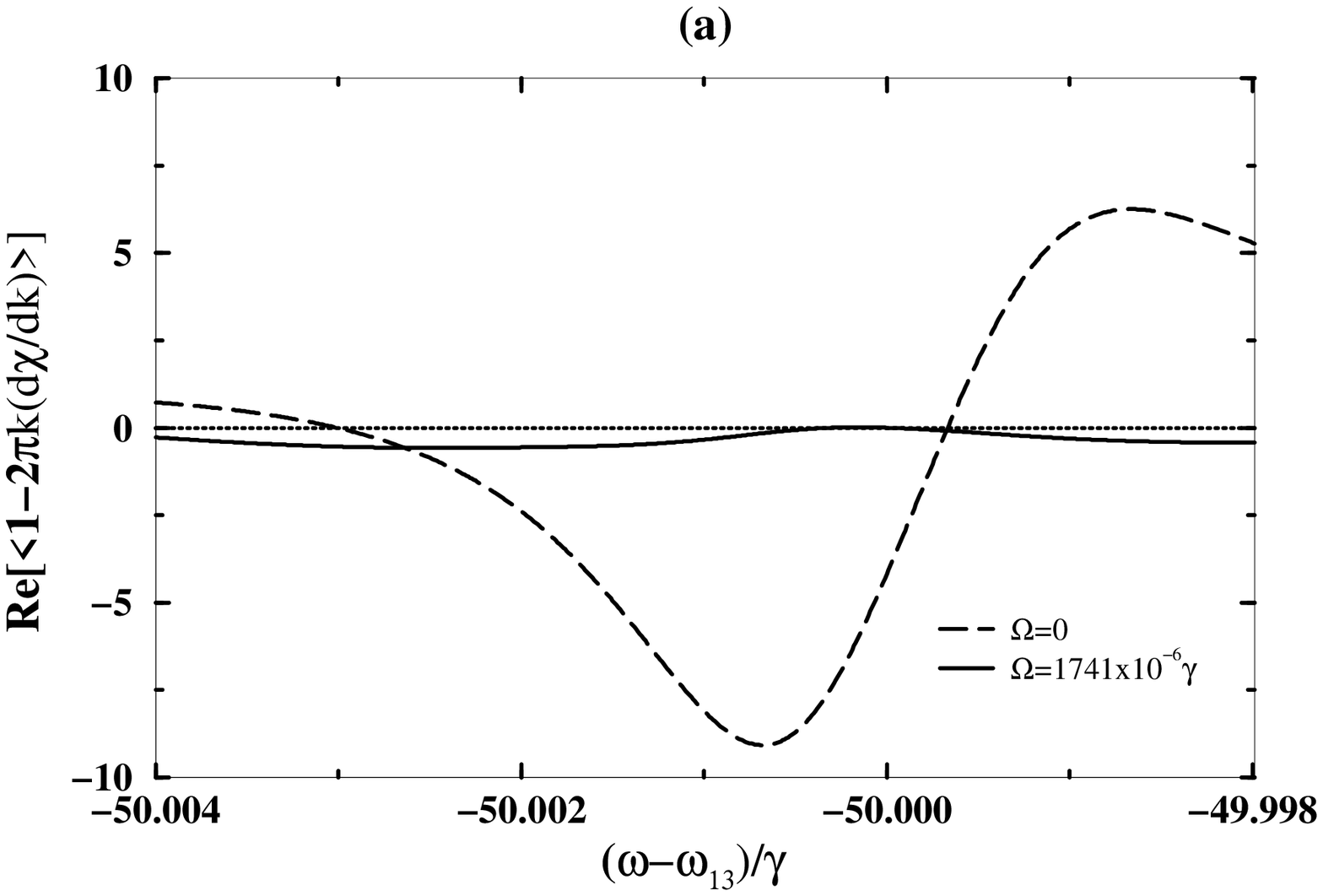,height=7.5cm,width=9.5cm}&
\psfig{figure=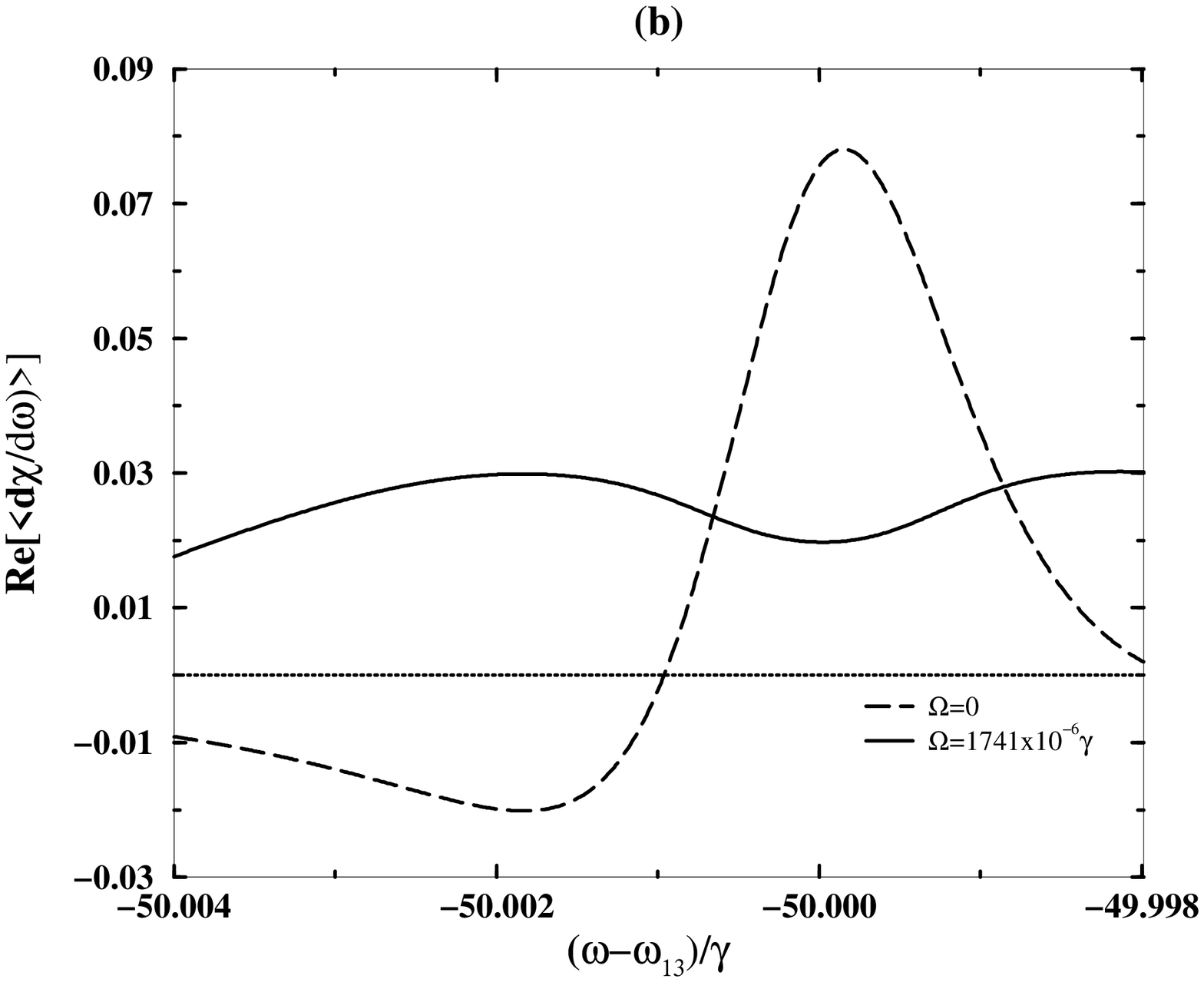,height=7.5cm,width=9.5cm}
\end{tabular}}
\caption{(a) shows the variation of the numerator in Eq.(2) with the probe
detuning taking Doppler effect into account. The Fig. 3(b) gives the slope
of the susceptibility.
The common parameters of the above two graphs for $^{87}$Rb vapor are chosen as:
Doppler width parameter D$=1.33\times10^9$ rad/sec, density
$n=10^{12}$ atoms/cc, $G=0.3\gamma$, $\Delta_3=0$,$\Delta_2=-50\gamma$
$\Gamma_{12}=\Gamma_{13}=0$, $\Gamma_{23}=0.001\gamma$,
$\gamma=3\pi\times10^{6}$~rad/sec. }
\end{figure}


\begin{thebibliography}{99}
\bibitem{footnote} It is not necessary to calculate upto first order in probe;
exact calculation is possible. However one studies concepts like group velocity
in a linear medium and therefore we treat probe to first order so that the
medium is linear as far as probe is concerned.
\end{thebibliography}
\end{document}